\documentstyle[12pt]{article}
\font\cmss=cmss10 \font\cmsss=cmss10 at 7pt
\def\IZ{\relax\ifmmode\mathchoice
{\hbox{\cmss Z\kern-.4em Z}}{\hbox{\cmss Z\kern-.4em Z}}
{\lower.9pt\hbox{\cmsss Z\kern-.4em Z}}
{\lower1.2pt\hbox{\cmsss Z\kern-.4em Z}}\else{\cmss Z\kern-.4em Z}\fi}

\def\@citex[#1]#2{%
\if@filesw \immediate \write \@auxout {\string \citation {#2}}\fi
\@tempcntb\m@ne \let\@h@ld\relax \def\@citea{}%
\@cite{%
  \@for \@citeb:=#2\do {%
    \@ifundefined {b@\@citeb}%
      {\@h@ld\@citea\@tempcntb\m@ne{\bf ?}%
      \@warning {Citation `\@citeb ' on page \thepage \space undefined}}%
      {\@tempcnta\@tempcntb \advance\@tempcnta\@ne%
      \@tempcntb\number\csname b@\@citeb \endcsname \relax%
      \ifnum\@tempcnta=\@tempcntb %   Number follows previous--hold on to it
        \ifx\@h@ld\relax%
          \edef \@h@ld{\@citea\csname b@\@citeb\endcsname}%
        \else%
          \edef\@h@ld{\ifmmode{-}\else--\fi\csname b@\@citeb\endcsname}%
        \fi%
      \else%   %  non-successor--dump what's held and do this one
        \@h@ld\@citea\csname b@\@citeb \endcsname%
        \let\@h@ld\relax%
      \fi}%
    \def\@citea{,\penalty\@highpenalty\,}%
  }\@h@ld
}{#1}}
\newcount\hour
\newcount\minute
\newtoks\amorpm
\hour=\time\divide\hour by 60
\minute=\time{\multiply\hour by 60 \global\advance\minute by-\hour}
\edef\standardtime{{\ifnum\hour<12 \global\amorpm={am}%
        \else\global\amorpm={pm}\advance\hour by-12 \fi
        \ifnum\hour=0 \hour=12 \fi
        \number\hour:\ifnum\minute<10 0\fi\number\minute\the\amorpm}}
\edef\militarytime{\number\hour:\ifnum\minute<10 0\fi\number\minute}

\def\draftlabel#1{{\@bsphack\if@filesw {\let\thepage\relax
   \xdef\@gtempa{\write\@auxout{\string
      \newlabel{#1}{{\@currentlabel}{\thepage}}}}}\@gtempa
   \if@nobreak \ifvmode\nobreak\fi\fi\fi\@esphack}
        \gdef\@eqnlabel{#1}}
\def\@eqnlabel{}
\def\@vacuum{}
\def\marginnote#1{}
\def\draftmarginnote#1{\marginpar{\raggedright\scriptsize\tt#1}}

\def\draft{\oddsidemargin -.5truein
        \def\@oddhead{\sl \phantom{\today\quad\militarytime} \hfil
        \smash{\Large\sl DRAFT} \hfil \today\quad\militarytime}
        \let\@evenhead\@oddhead
        \let\label=\draftlabel
        \let\marginnote=\draftmarginnote
        \def\ps@empty{\let\@mkboth\@gobbletwo
        \def\@oddfoot{\hfil \smash{\Large\sl DRAFT} \hfil}
        \let\@evenfoot\@oddhead}
        \def\@eqnnum{(\theequation)\rlap{\kern\marginparsep\tt\@eqnlabel}%
        \global\let\@eqnlabel\@vacuum}  }

\def\nblack{            % For people without blackboard fonts
\def\ZZ{Z \n{10} Z}
\def\NN{N \n{14} N}
\def\CC{C \n{11} C}
\def\RR{R \n{11} R}
\def\QQ{Q \n{12} Q}
\def\PP{P \n{11} P}
}

\def\opeq#1{\advance\lineskip#1 \advance\baselineskip#1
        \advance\lineskiplimit#1}
\def\eqalign#1{\null\,\vcenter{\opeq{2.5\jot}\mathsurround=0pt
        \everycr={}\tabskip=0pt
        \halign{\strut\hfil$\displaystyle{##}$&$\displaystyle{{}##}$\hfil
        \crcr#1\crcr}}\,\null}

\def\eqalignno#1{\displ@y \tabskip\centering
  \halign to\displaywidth{\hfil$\@lign\displaystyle{##}$\tabskip\z@skip
    &$\@lign\displaystyle{{}##}$\hfil\tabskip\centering
    &\llap{$\@lign##$}\tabskip\z@skip\crcr
    #1\crcr}}

\catcode`\@=12

\def\beq{\begin{equation}}
\def\eeq{\end{equation}}
\def\beqar{\begin{eqnarray}}
\def\eeqar{\end{eqnarray}}

\def\nfrac#1#2{{\displaystyle{\vphantom1\smash{\lower.5ex\hbox{\small$#1$}}%
        \over\vphantom1\smash{\raise.25ex\hbox{\small$#2$}}}}}

\def\p#1{\mskip#1mu}
\def\n#1{\mskip-#1mu}
\def\stop{\p6.}
\def\comma{\p6,}

\def\noj#1,#2,{{\bf #1} (19#2)\ }
\def\jou#1,#2,#3,{{\sl #1\/ }{\bf #2} (19#3)\ }
\def\ann#1,#2,{{\sl Ann.\ Physics\/ }{\bf #1} (19#2)\ }
\def\cmp#1,#2,{{\sl Comm.\ Math.\ Phys.\/ }{\bf #1} (19#2)\ }
\def\cq#1,#2,{{\sl Class.\ Quantum Grav.\/ }{\bf #1} (19#2)\ }
\def\cqg#1,#2,{{\sl Class.\ Quantum Grav.\/ }{\bf #1} (19#2)\ }
\def\ijmp#1,#2,{{\sl Int.\ J.\ Mod.\ Phys.\/ }{\bf A#1} (19#2)\ }
\def\jmp#1,#2,{{\sl J.\ Math.\ Phys.\/ }{\bf #1} (19#2)\ }
\def\lmp#1,#2,{{\sl Lett.\ Math.\ Phys.\/ }{\bf #1} (19#2)\ }
\def\grg#1,#2,{{\sl Gen.\ Rel.\ Grav.\/ }{\bf #1} (19#2)\ }
\def\mpl#1,#2,{{\sl Mod.\ Phys.\ Lett.\/ }{\bf A#1} (19#2)\ }
\def\nc#1,#2,{{\sl Nuovo Cim.\/ }{\bf #1} (19#2)\ }
\def\np#1,#2,{{\sl Nucl.\ Phys.\/ }{\bf B#1} (19#2)\ }
\def\pl#1,#2,{{\sl Phys.\ Lett.\/ }{\bf #1B} (19#2)\ }
\def\pla#1,#2,{{\sl Phys.\ Lett.\/ }{\bf #1A} (19#2)\ }
\def\pr#1,#2,{{\sl Phys.\ Rev.\/ }{\bf #1} (19#2)\ }
\def\prd#1,#2,{{\sl Phys.\ Rev.\/ }{\bf D#1} (19#2)\ }
\def\prl#1,#2,{{\sl Phys.\ Rev.\ Lett.\/ }{\bf #1} (19#2)\ }
\def\prp#1,#2,{{\sl Phys.\ Rept.\/ }{\bf #1C} (19#2)\ }
\def\ptp#1,#2,{{\sl Prog.\ Theor.\ Phys.\/ }{\bf #1} (19#2)\ }
\def\ptpsup#1,#2,{{\sl Prog.\ Theor.\ Phys.\/ Suppl.\/ }{\bf #1} (19#2)\ }
\def\rmp#1,#2,{{\sl Rev.\ Mod.\ Phys.\/ }{\bf #1} (19#2)\ }
\def\yadfiz#1,#2,#3[#4,#5]{{\sl Yad.\ Fiz.\/ }{\bf #1} (19#2) #3%
\ [{\sl Sov.\ J.\ Nucl.\ Phys.\/ }{\bf #4} (19#2) #5]}
\def\zh#1,#2,#3[#4,#5]{{\sl Zh.\ Exp.\ Theor.\ Fiz.\/ }{\bf #1} (19#2) #3%
\ [{\sl Sov.\ Phys.\ JETP\/ }{\bf #4} (19#2) #5]}

\def\ie{\hbox{\it i.e.\/}}

\def\to{\rightarrow}

\def\lae{\mathrel{\mathop{\smash{\lower .5 ex \hbox{$\stackrel<\sim$}}}}}
\def\lae{\mathrel{\mathop{\smash{\lower .5 ex \hbox{$\stackrel>\sim$}}}}}

\def\pa{\partial}

\def\l:{\mathopen{:}\,}
\def\r:{\,\mathclose{:}}
\def\[{\left[}          \def\]{\right]}
\def\({\left(}          \def\){\right)}
\def\<{\left<}          \def\>{\right>}

\def\CT{{\cal T}}
\def\CY{{\cal Y}}
\def\res{{\rm res}}

\nblack

\catcode`\@=11
\def\theequation{\thesection.\arabic{equation}}
\@addtoreset{equation}{section}
\@addtoreset{footnote}{section}
\@addtoreset{footnote}{subsection}
\catcode`\@=12
\begin{document}
\begin{titlepage}

\begin{center}
Jan 6, 1994\hfill      IASSNS-HEP-94/1 \\
\hfill                  TAUP--2130--93 \\
\hfill                  WIS-93/123/Dec-PH               \\
\hfill                  hep-th/9401030

\vskip 1 cm

{\large \bf
Topological Landau-Ginzburg Formulation and Integrable Structure
of 2d String Theory\\
}
\vskip .1 cm

\vskip .5 cm
{Amihay Hanany\footnote{
E-Mail: ftami@weizmann.weizmann.ac.il}}
\vskip 0.2cm

{\sl
Department of Particle Physics \\
Weizmann Institute of Science \\
 76100 Rehovot Israel
}

{Yaron Oz\footnote{
Work supported in part by the US-Israel Binational Science Foundation,
and the Israel Academy of Science.  E-Mail:
yarono@ccsg.tau.ac.il}
}
\vskip 0.2cm

{\sl
School of Physics and Astronomy\\Raymond and Beverly Sackler Faculty
of Exact Sciences\\Tel-Aviv University\\Ramat Aviv, Tel-Aviv 69978, Israel.
}

{M. Ronen Plesser\footnote{
Work supported in part by the National Science Foundation under grant
NSF \#PHY92-45317 and by the W. M. Keck Foundation.
E-Mail: plesser@guinness.ias.edu}}
\vskip 0.2cm

{\sl
School of Natural Sciences
\\ Institute for Advanced Study
\\ Olden Lane
\\ Princeton, NJ 08540
}
\end{center}

\vskip .5 cm
\newpage
\begin{abstract}
We construct a topological Landau-Ginzburg formulation of the two-dimensional
string at the self-dual radius. The model is an analytic continuation of the
$A_{k+1}$ minimal model to $k=-3$.
We compute the superpotential and calculate tachyon correlators in the
Landau-Ginzburg framework. The results are in complete agreement with
matrix model calculations.
We identify the momentum one tachyon as the puncture operator,
non-negative momentum tachyons as primary fields, and negative momentum
ones as descendants.
The model thus has an infinite number of primary fields,
and the topological metric vanishes on the small phase space
when restricted to these.
We find a parity invariant multi-contact algebra
with irreducible contact terms of arbitrarily large number of fields.
The formulation of this Landau-Ginzburg description in terms of period
integral coincides with the genus zero $W_{1+\infty}$ identities
of two-dimensional string theory.
We study the underlying Toda lattice integrable hierarchy in the Lax
formulation and find that the Landau-Ginzburg superpotential coincides
with a derivative of the Baker-Akhiezer wave function in the
dispersionless limit. This establishes a connection between the
topological and integrable structures.
Guided by this connection we derive relations formally analogous to
the string equation.

\end{abstract}

\end{titlepage}

\section{Introduction}
In recent years much attention has been devoted to string theories
that are solvable by matrix model techniques, in the hope that
insights into the fundamental nature of string theory could be
gleaned from their solutions. Probably the most promising of these
is the theory of $c=1$ conformal matter coupled to two-dimensional
gravity. This theory describes string dynamics in a two-dimensional
spacetime, with one propagating field-theoretic degree of freedom -
the massless scalar ``tachyon'', as well as so-called special states
\cite{Lian,Mukherji,Bouk}
at discrete values of (Euclidean) momentum, and is probably related to
the two-dimensional black hole of \cite{witbh}. Tachyon dynamics in
this model have been extensively studied by matrix model techniques
\cite{twodreviews} as well as continuum methods \cite{kdf},
and the full scattering matrix computed
\cite{exacts}. The dynamics of the special states is not well
understood. Their manifestation in the matrix model is not known and
continuum studies of some of their bulk correlators have
yielded divergent results, which vanish upon regularization
\cite{Li}.

One of the most striking successes in the study of the unitary minimal
models coupled to gravity was the discovery of an underlying
integrable structure \cite{Douglas,migros,dvv,fukuma} as well as
a topological phase \cite{WittTG,KekeLi,KekeLi2}, and the intimate
relation between these \cite{Wittinter,Wittalgeom,Kosevi}. The
analogous connection in the $c=1$ case has not been fully clarified,
and one may hope that when understood this connection will lead to a
better understanding of the model.

In the parameter space of two-dimensional string theory there is a
natural point at which to seek these structures. In a Euclidean
spacetime one introduces the radius of the compact spatial slice, and
the integrable structure appears most naturally when this is set to
the self-dual value. Tachyon scattering amplitudes can be analytically
continued to this value of the radius (note we ignore a possible phase
transition and deal with the amplitudes continued from large radius;
in particular we will not be introducing winding modes anywhere in
this work) using the results of \cite{Kleblowe}. This was pursued in
\cite{ronen} and the integrable structure that emerges is
the Toda lattice hierarchy; the generating function for tachyon
correlators is a tau function for this hierarchy. This leads to a set
of $W_{1+\infty}$ constraints on the amplitudes which
completely determine the scattering matrix.\footnote{This same algebra has
appeared in other studies of the same model
\cite{Mandal,Wittds,Poly,KMS,Kitazawa,kleb,sakai,Zwie} and the
relations between these appearances is unclear.} These will be crucial
to our study and will be discussed in section 2. For the remainder of
the paper we restrict attention to this point in parameter space.

Witten \cite{WittenNM} discovered that the partition function of
the $N=2$ twisted
minimal model realized as an $SU(2)/U(1)$ coset at level $k = -3$
(more properly thought of as $SL(2)/U(1)$ at level 3)
computes the Euler characteristic of the moduli space of Riemann
surfaces  (which is known to be the partition function
of the $c=1$ system \cite{ecgk,eulerchar}).
This led to the conjecture that $A_{k+1}$ theory at $k = -3$
corresponds to the $2d$ string at the self-dual radius.
Further support to the conjecture was given by Vafa and Mukhi
\cite{Vafa} who showed that appropriate choices of the  $SL(2)$
representations reproduced the spectrum of states of $2d$ strings, and
argued at the level of Lagrangians that the two models are
equivalent.\footnote{The spectrum of the $SU(2)/U(1)$ coset at $k=-3$ was
shown to be in $1-1$ correspondence with the spectrum of the $2d$
string using a
unitary transformation between the corresponding Hilbert spaces
\cite{G/G} .} These authors also
observed that the  four-point function computed by intersection theory
in \cite{WittenNM} yields the correct tachyon four-point function when
analytically continued to $k=-3$. Subsequently,
$1\rightarrow n$ tachyon amplitudes, i.e. one
negative momentum and $n$ positive momentum tachyons or vice versa,
as well as 5 point function in various
kinematic regions have been calculated in the same manner showing complete
agreement with matrix model and continuum Liouville calculations
\cite{Mooreron}.
A complete understanding of the topological field
theory, similar to that accomplished in the $k>0$ case by relating it
to the integrable structure, has not been achieved.

This work has brought up some differences between the topological
version of two dimensional string theory and the analogous object for
the minimal unitary models. One difference is that the number of
primary fields appears to be infinite. This means that the recursion
relations of topological gravity must be treated with care to avoid
divergent answers. Another is the nature of the factorization
relations. The $W_{1+\infty}$ recursion relations of \cite{ronen} are
highly nonlinear, in contrast to the quadratic nature of the usual
recursion relations. Some of these properties may be shared by other
potentially interesting topological field theories (such as those
related to topological Yang-Mills theory in two dimensions
and topological sigma models with noncompact target spaces) and
understanding the
nature of these more general topological field theories provides added
motivation for studying this model.
Finally, the topological field theory in this
case seems to compute directly the scattering amplitudes of the $c=1$
system coupled to Liouville gravity; there is no phase transition
separating the two, and at genus zero the amplitudes are numerically
identical.

In this paper we construct a topological
Landau-Ginzburg theory which computes the tachyon scattering
amplitudes. The model is described by the
superpotential $W={1\over k+2} x^{k+2}$ with $k=-3$. We do not derive
the amplitudes from a well-defined worldsheet action, but
extend formally the rules for computing
correlators in Landau-Ginzburg theories to define our model. We will,
however, present what we think are very strong arguments that this is
the correct formulation of the model, in addition to the suggestive
speculations about analytic continuation in $k$. These include
computations showing that the correlators of two-dimensional
string theory are indeed reproduced, as well as a relation, extending
that found for $k>0$, between the Landau-Ginzburg superpotential and
the Lax operator of the related integrable hierarchy.

The paper is organized as follows. In section two we construct the
topological Landau-Ginzburg model by extrapolation from a few
explicitly computed amplitudes. We study some of the properties of the
model; in particular we are led to an identification of the primary
fields as the tachyons with non-negative momenta.
Tachyons with negative momenta are identified with gravitational
descendants. We continue this
study in section three, restricting attention to the computation of
arbitrary correlators (of primaries and descendants) on the ``small
phase space'' \cite{Dijwit}. We find a rich algebra of contact terms
which contains infinitely many irreducible terms, and use this to
verify that our Landau-Ginzburg model reproduces tachyon scattering
amplitudes for some infinite families of correlators. In section four
we reformulate the model following \cite{EguchiP}; we show how the
$W_{1+\infty}$ recursion relations arise naturally in this formulation.
In section five we forge the connection to the Toda lattice hierarchy.
In particular, we show that the superpotential we have constructed is
a derivative of the (semiclassical) Baker-Akhiezer function, in
analogy with the relation to the Lax operator in minimal models
\cite{DVV}. We identify some relations analogous to the string
equation and speculate on the all-genus version of our results.
Section six summarizes our conclusions, and some detailed calculations
are relegated to the appendix.

\section{Topological Landau-Ginzburg description of 2d string theory}
In this section we propose a topological Landau-Ginzburg formulation
for two-dimensional string theory. We begin by reviewing the structure
of the tachyon correlators at the self-dual radius
which we will reproduce using Landau-Ginzburg techniques. We compute
the first few amplitudes explicitly.
Then we construct a Landau-Ginzburg model
which reproduces these correlators. Finally, extrapolating these
results, we define our proposed Landau-Ginzburg model in generality
and show that it produces a well-defined set of correlators. We
conjecture that these will be identical to those of the $2d$ string.
At the self-dual radius the tachyon operators are
labelled by integers representing the momentum of the tachyon state
they create. We will use the notation
\begin{equation}
\<\<{\cal O}\>\> = \< {\cal O}e^{\sum_{n=-\infty}^\infty t_n \CT_n} \>
\label{defcor}
\end{equation}
where the right-hand side is defined as a function of $t$ by a formal
series expansion, and $\cal O$ is any product of tachyon operators.

\subsection{$W_{1+\infty}$ recursion relations of 2d string theory}
The tachyon correlators obey a set of $W_{1+\infty}$ recursion
relations which  completely determine them \cite{ronen}. Restricting
attention to genus zero, these are
\begin{equation}
\<\<\CT_{n}\>\> \equiv \<\CT_{n}\exp\[\sum_{k=-\infty}^{\infty}t_k \CT_k\]\>
= \frac{1}{n(n+1)}\res(-\bar{W})^{n+1}\comma
\label{pwi}
\end{equation}
for $n>0$, where
\begin{equation}
\bar W = -\frac{1}{x} \[1 + \sum_{k > 0} t_{-k} x^k +
\sum_{k > 0} k x^{-k}\<\<\CT_{-k}\>\>\]
\comma
\label{barW}
\end{equation}
and $\res$ denotes the residue at infinity; this picks out
of a Laurent series in $x$ the coefficient of $x^{-1}$.
The correlators are symmetric under a parity transformation that takes
$\CT_n$ to $\CT_{-n}$. Thus
\begin{equation}
\<\<\CT_{-n}\>\> \equiv \<\CT_{-n}\exp\[\sum_{k=-\infty}^{\infty}t_k \CT_k\]\>
= \frac{1}{n(n+1)} \res(-W)^{n+1}\comma
\label{nwi}
\end{equation}
for $n>0$, with
\begin{equation}
W = -\frac{1}{x} \[1 + \sum_{k > 0} t_k x^k +
\sum_{k > 0} k x^{-k}\<\<\CT_{k}\>\>\]
\stop
\label{W}
\end{equation}
The symmetry under parity asserts that $W(\{t_k\}) = \bar W(\{t_{-k}\})$.

The relations (\ref{pwi}) can be explicitly written out \cite{ronmoore}
\begin{equation}
\eqalign{
\<\CT_n\prod_{i = 1}^m \CT_{n_i}\> &= \sum_{k = 2}^{min(m, n+1)}
\frac{\Gamma(n)}{\Gamma(2+n-k)}\cr
\sum_{l = 1}^{min(m_-, k-1)}&\sum_{|T| = l}\sum_{S_1,\ldots,S_{k-l}}
\prod_{j=1}^{k-l}\[\theta(n(S_j))n(S_j)\<\CT_{-n(S_j)}\prod_{S_j}\CT_{n_i}\>\]
\ .\cr
}
\label{MPWI}
\end{equation}
The notation used in (\ref{MPWI}) is as follows : Let
$S=\{n_1,\dots,n_m \}$, $S_-$ denotes the subset of $S$ of negative momenta.
Denote $m_- = |S_-|$. The sum on $T$ is over subsets of $S_-$ of order $l$.
The subsequent sum is over distinct disjoint decompositions
$S_1\cup\cdots\cup S_{k-l} = S \setminus T$. $n(T)$ denotes the sum of momenta
in the set $T$, and $\theta(x)$ is the Heaviside function.

The most striking feature of these relations, which we will need to
reproduce in our Landau-Ginzburg formulation, is that they are highly
nonlinear. This nonlinearity is actually twofold. The first form of
nonlinearity is related to the fact that $T_n$ ``absorbs'' $l$ other
tachyons where $l$ can be as large as $n$. This has been much
discussed \cite{KMS,kleb,Zwie}. Even more strikingly,
the right-hand side of (\ref{MPWI}) contains products of up
to $n+1-l$ correlators. In the Landau-Ginzburg theory this will manifest
itself as the existence of an infinite set of irreducible contact
terms of $n$ operators for arbitrarily large $n$. A worldsheet
interpretation of these relations (which are in some sense the
$k\to\infty$ limit of the $W_{k+2}$ recursion relations of \cite{dvv,fukuma}
with the difference that the $W$ currents do not annihilate the
partition function but generate tachyon insertions) would thus need to
involve nonzero contributions from high-codimension boundaries of
moduli space \cite{ronmoore}.

The correlators $\<\prod_{j=1}^s \CT_{n_j}\>$
determined by these relations have the following
properties as functions of the momenta $n_j$ \cite{exacts}. We
decompose the set of
momenta such that $\sum n_j = 0$ into cones separated by hyperplanes
on which any partial sum of momenta vanishes. On any one of these
cones (determining various kinematic regimes) the amplitude is given
by a polynomial in the $n_j$ of order $s-3$, with integer
coefficients.
At the hypersurfaces bounding cones, the coefficients of this
polynomial jump so that the amplitudes are continuous.

We can explicitly write the first few correlators, exemplifying these
properties:
\beq
\<\CT_n\CT_{-n}\> = \frac{1}{|n|}\comma
\label{2pt}
\eeq

\beq
\<\CT_n\CT_{n_1}\CT_{n_2}\> =  1 \comma
\label{3pt}
\eeq

\beq
\<\CT_n\CT_{n_1}\CT_{n_2}\CT_{n_3}\> =
(n-1) - \sum_{i=1}^3 (n + n_i)\theta(-n - n_i)
\comma
\label{4pt}
\eeq

\begin{equation}
\eqalign{
\<\CT_n\prod_{i=1}^4\CT_{n_i}\> &=  (n-1)(n-2)
-(n-1)\sum_{i,j = 1, i \neq j}^4(n+n_i+n_j) \theta(-n - n_i - n_j)\cr
&\qquad - \sum_{i=1}^4 (n + n_i)
\theta(-n - n_i)\<\CT_{n + n_i}\prod_{i \neq j=1}^4\CT_{n_j}\>
\ ,\cr
}
\label{5pt}
\end{equation}
where we have factored out the $\delta$ function enforcing momentum
conservation.

Of particular interest for the sequel is the $\CT_1$ identity
of (\ref{MPWI})
\beq
\<\CT_1\prod_{i = 1}^m \CT_{n_i}\> =
- \sum_{i = 1}^m (n_i+1)\theta( - n_i-1)
\<\CT_{n_i+1}\prod_{j \neq i}^m \CT_{n_j}\>\stop
\label{T1WI}
\eeq
The similarity of this to the puncture equation in topological gravity
has been remarked in \cite{Kitazawa}.

\subsection{Construction of the Landau-Ginzburg superpotential}

Our aim is to construct a Landau-Ginzburg superpotential $W$ from which the
tachyon correlators of the previous section are derived. By this we mean
precisely the following:
\begin{equation}
\<\<\CT_i\CT_j\CT_k\>\> = \res\(\frac{\phi_i(x,t)\phi_j(x,t)\phi_k(x,t)}
{\partial_x W(x,t)}\) \comma
\label{3}
\end{equation}
with $\phi_n = - \partial_n W$, and
\beq
\<\prod_{i=1}^m\CT_{n_i}\> =
\pa_{n_m}\cdots\pa_{n_4}\<\<\CT_{n_1}\CT_{n_2}\CT_{n_3}\>\>(t=0)
\stop
\label{mpt}
\eeq
We note that the second of these equations is of course true only
in a particular set of coordinates on the parameter space of the
theory; thus in finding our solution we will have also found the
``flat coordinates'' in which (\ref{mpt}) holds without additional
contact terms. As we point out later, the definition of these
coordinates in terms of the metric on primary fields is not useful
for the model at hand, but the existence of preferred coordinates
in which contact terms vanish is a general feature of topological
theories and continues to hold here.

The candidate topological Landau Ginzburg description of the $2d$
string is the analytic continuation to $k=-3$ of
the $A_{k+1}$ minimal theories, \ie\ a superpotential given by
\beq
W(x)=\frac{1}{k+2}x^{k+2}\stop
\label{ALG}
\eeq
We want to represent the tachyon fields $\phi_k$ as functions of $x$.
At $t=0$ we are led by momentum conservation of the tachyon three-point
function and the form of (\ref{3}) to
\begin{equation}
\phi_k(t=0) = x^{k-1}
\label{phi0}
\end{equation}
so that
\begin{equation}
W = -\frac{1}{x}\(1 + \sum_{n=-\infty}^\infty t_n x^n\) + {\cal O}(t^2)\ .
\label{W0}
\end{equation}

Comparing (\ref{mpt}) with (\ref{4pt}) we find the correction to
$\phi$ linear in $t$
\beq
\phi_n = x^{n-1}\[1 - \sum_{m=-\infty}^{\infty}\psi(n + m)t_mx^m\]+
O(t^2)\comma
\label{phit}
\eeq
where $\psi(x) = x\theta(-x)$.
This, in turn, can be integrated to give the superpotential to second
order in $t$. The $O(t^2)$ correction to $\phi_n$ is found by matching
to the tachyon five point correlator as
\begin{equation}
\phi_n = x^{n-1}\[1 - \sum_{m=-\infty}^{\infty}\psi(n + m)t_mx^m -
\frac{1}{2}\sum_{m,l=-\infty}^\infty \psi(n+m+l) t_m t_l x^{m+l}
\<\CT_n\CT_m\CT_l\CT_{-n-m-l}\>\]\ .
\end{equation}

\subsection{The topological LG model}

In the previous subsection we have constructed, order by order in $t$, a
superpotential $W$ which reproduces the tachyon scattering amplitudes
via (\ref{mpt}). Extrapolating the results of these steps in a natural way,
one finds the superpotential
\begin{equation}
W = -\frac{1}{x}\[1 + \sum_{k=-\infty}^{\infty}t_kx^k -
\sum_{k > 1}\frac{1}{k!} \sum_{n_1,\ldots,n_k = -\infty}^{\infty}
\psi(\Sigma n_i)
\prod_{i=1}^k
t_{n_i}x^{n_i}\<\CT_{-\Sigma n_i}\prod_{i=1}^k\CT_{n_i}\>\] \stop
\label{itW}
\end{equation}

We now propose the following as the Landau-Ginzburg description of
$2d$ string theory. The superpotential is given by (\ref{itW}), the
correlators by (\ref{mpt}) with $\phi_n = -\partial_n W$. Since these
same correlators are the coefficients of $W$ we view (\ref{mpt}) as a
recursion relation which determines the $m$-point function in terms of
lower orders. Thus this definition recursively fixes a unique set
of correlators. We claim that this is identical to the
tachyon scattering matrix. We do not directly prove this but provide
some additional evidence in the sequel.
Note that this statement implies that the amplitudes
exhibit some properties not manifest in this formulation. In particular,
they must be symmetric under permutation of the operators, as well as
under the parity transformation mentioned above.

Let us study the model we propose a bit more closely. The first question
one must answer is what are the primary operators. We suggest that these
are the tachyons of non-negative momentum. The negative-momentum tachyons
are thus interpreted as gravitational descendants. Their expression in terms
of $x$ (we have not introduced ghosts) follows the ideas of
\cite{Lossev,EguchiD} for the case $k>0$. We base this suggestion on the
following:

-- With this interpretation, (\ref{T1WI}) becomes simply the
puncture equation, with $\CT_1$ the puncture operator.

-- In the $A_{k+1}$ model ($k>0$), the ring of primary fields
is the ring of polynomials in $x$
modulo the ideal generated by $\pa_x W(x)$.
At $k=-3$ the ideal does not exist and we are led to consider
the whole ring of polynomials as the ring of primary fields.

With this interpretation (\ref{itW}) gives the dependence of the
superpotential on the times coupling to primary {\it and descendant}
states. This is in contrast to the usual study of Landau-Ginzburg
theories which restricts attention to the ``small phase space'' where
the couplings of descendant fields vanish. This is important in this
theory because momentum conservation constrains the correlators of
primary operators to vanish, so descendants must be introduced to
obtain nontrivial results.

Equation (\ref{itW}) can be rewritten in a nicer form as:
\beq
W = -\frac{1}{x} \[1 + \sum_{k > 0} t_k x^k +
\sum_{k > 0} k x^{-k}\<\<\CT_{k}\>\>\]
\stop
\label{SW}
\end{equation}
We already encountered this object in the $2d$ string $W_{1+\infty}$
identities (see equation (\ref{W})).
This is no surprise as we will see in the next sections.
The Landau-Ginzburg fields corresponding to the tachyons are:
\beq
\phi_n = \theta(n)x^{n-1} + \sum_{m > 0} m x^{-m-1}\<\<\CT_m\CT_n\>\>\stop
\label{Phi}
\eeq

Of course, one can alternatively write $\bar\phi_n =x^{-n-1}+{\cal O}(t)$
and proceed along the same lines as above. One ends up with $\bar{W}$
of equation (\ref{barW}) as the superpotential and
\beq
\bar\phi_n = \theta(-n)x^{-n-1} + \sum_{m > 0} m
 x^{-m-1}\<\<\CT_{-m}\CT_n\>\>\comma
\label{baPhi}
\eeq
as the Landau-Ginzburg fields. That the two descriptions
are completely equivalent reflects the parity invariance of the theory.

\section{Contact algebra}
In this section we will work in the small phase space, \ie\ only the
gravitational primaries are turned on. In light of our identification
of the non-negative momentum tachyons as primaries, we turn on the coupling
to them. The picture that will emerge closely resembles
that of the minimal topological theories. However, unlike the minimal models
in which the study of correlators of only primaries is meaningful, here
this leads to an empty structure as all the correlators vanish by momentum
conservation. Thus we will be studying the correlators of primary
and descendant operators on the small phase space. We will find a
set of contact terms that reflect the nonlinearity of the
recursion relations to which we alluded earlier, and use these to
demonstrate for a subset of the correlators
that the Landau-Ginzburg theory described in the
previous section indeed satisfies the $W_{1+\infty}$ constraints.
We believe it will be possible to
prove along these lines that the two sets of correlators are
completely identical.
This section will contain some detailed calculations; these serve
to demonstrate explicitly the claims made above that the form we
assume for the superpotential determines the correlators
unambiguously.
To avoid cumbersome notation we will assume that in any
calculation we set $t_n = 0$ for $n<0$ {\it after} all derivatives
have been taken.

On the small phase space, the superpotential (\ref{SW}) takes the form
\beq
W=-{1\over x}\(1+\sum_{k=0}^\infty t_kx^k\) \stop
\label{WPS}
\eeq
Note that $W$ is linear in $t$, reflecting the absence of contact terms
between primary operators in this theory.
Indeed $\phi_n = -\partial_n W = x^{n-1}$ for $n>0$ is $t$-independent.

We can now compute $\phi_{-n} =
\sum_{m>0} m x^{-m-1} \<\<\CT_m\CT_{-n}\>\>$ (from (\ref{Phi})).
To do so most easily, note that of course $\phi_{-n}(t=0) =
x^{-n-1}$.
Also,
\begin{equation}
\eqalign{
\partial_k\phi_{-n} &=
\sum_{m>0} m x^{-m-1} \res \left({\phi_m\phi_k\phi_{-n}\over
W'}\right)\cr
 &= -\partial_x\left({x^{k-1}\phi_{-n}\over W'}\right)_- \ ,
}
\label{phimi}
\end{equation}
where $f(x)_-$ extracts the negative powers of $x$ in the Laurent series
$f$ and primes denote derivatives with respect to $x$.
These are a set of first order equations for the $t$ dependence of
$\phi_{-n}$; their integrability can be checked easily but
we will simply demonstrate an explicit solution:
\beq
\phi_{-n}=-{1\over n}\[(-W)^n\]^\prime_- \comma
\label{phips}
\eeq
The representation (\ref{phips}) of $\phi_{-n}$ in terms of the
superpotential parallels the description of the primary fields in
the minimal models \cite{DVV}:
\beq
\phi_{n-1} = \frac{1}{n}\[ ((k+2)W)^{n\over k+2}\]^\prime_+
\label{phimin}
\eeq
where $f(x)_+ = f(x)-f(x)_-$ extracts the nonnegative powers.
We can use (\ref{phimin}) in (\ref{3}) to see that the result of
\cite{kdf} for
the $1\to n$ amplitudes is reproduced; in terms of the notation we
are using we have shown that (\ref{nwi}) holds
on the small phase space.

We can continue, using the same methods, to show that this still holds
order by order in the deviation from the small phase space. At the next
step we note that (\ref{nwi}) implies:
\begin{equation}
\<\<\CT_{-n_1}\CT_{-n_2}\>\> =
\res \(\frac{1}{n_1} (-W)^{n_1} \phi_{-n_2} \)
\label{twomin}
\end{equation}
by taking the derivative. On the other hand we can compute
\begin{equation}
\<\<\CT_k\CT_{-n_1}\CT_{-n_2}\>\> =
\res \left( {x^{k-1} \phi_{-n_1}\phi_{-n_2}\over W'}\right)
\end{equation}
with the help of a useful identity following from (\ref{phips})
\begin{equation}
\phi_{-n} = (-W)^{n-1} W' + \frac{1}{n}\[(-W)^n\]'_+
\label{useful}
\end{equation}
and some obvious properties of $\res$
\begin{equation}
\eqalign{
\<\<\CT_k\CT_{-n_1}\CT_{-n_2}\>\> &=
\res \left\{ \[x^{k-1}(-W)^{n_1-1}\] \phi_{-n_2} +
\[x^{k-1} (-W)^{n_2-1}\]_- \[\frac{1}{n_1}(-W)^{n_1}\]'\right\}\cr
&= \res\left\{
\phi_{-n_2} \partial_k\left(\frac{1}{n_1}(-W)^{n_1}\right)
+ \left(\frac{1}{n_1}(-W)^{n_1}\right) \partial_k \phi_{-n_2}
\right\}\cr
}
\label{twomini}
\end{equation}
where we used (\ref{phimi}). Integrating we obtain agreement with
(\ref{twomin}), extending (\ref{nwi}) to first order in the deviation.
In terms of correlators, we have shown that the Landau-Ginzburg description
yields the correlators of two-dimensional string theory for $2\to n$
amplitudes.

We have carried this computation through to the $4\to n$ amplitudes using
these methods. The interested reader can find the details in the appendix.
That the $4\to n$ amplitudes are computed correctly is important in that
these necessarily involve the dependence of $\phi_{-n}$ on the $t_{-m}$,
so their agreement with $2d$ string results implies the correctness of
our contact terms between descendants. We strongly believe that an
inductive proof to all orders
that (\ref{nwi}) is satisfied can be constructed along these lines.

In performing this calculation one obtains a set of contact terms
determining the dependence of $\phi_n$ on the times coupling to primary
{\it and descendant} states. For example, (\ref{phimi}) generalizes to
\beq
{\partial_n\phi_m=-\left({\phi_n\phi_m\over W^\prime}
\right)^\prime}_-  \comma
\label{cont}
\eeq
where $n,m$ are either positive or negative. Thus, we identify the contact
term associated with the collision of two operators as
\beq
C(\phi_n,\phi_m)=-\({\phi_n\phi_m\over W^\prime}\)^\prime_- \stop
\label{contact}
\eeq
The analogous contact in the minimal models reads
\cite{Lossev,LossevP,EguchiD}
\beq
C(\phi_n,\phi_m)=\({\phi_n\phi_m\over W^\prime}\)^\prime_+ \stop
\label{mincontact}
\eeq
Note, however, that while (\ref{contact}) holds both for primaries
(non-negative momentum tachyons) and descendants (negative momentum tachyons),
(\ref{mincontact}) holds when one of the operators is a primary.

One can proceed further to get multi-contacts, for instance:
\begin{equation}
\eqalign{
\partial_{n_1}\partial_{n_2}\phi_{n_3}=C(C(\phi_{n_1},\phi_{n_2}),
\phi_{n_3}) +
C(C(\phi_{n_1},\phi_{n_3}),\phi_{n_2}) &+
C(C(\phi_{n_2},\phi_{n_3}),\phi_{n_1})\cr
&+C(\phi_{n_1},\phi_{n_2},\phi_{n_3}) \ ,\cr
}
\label{contt}
\end{equation}
where
\beq
C(\phi_{n_1},\phi_{n_2},\phi_{n_3}) =
-\(\({\phi_{n_1}\phi_{n_2}\phi_{n_3}\over W^\prime}\)^\prime
{1\over W^\prime}\)^\prime_-  \stop
\label{multcont}
\eeq
The form of (\ref{multcont}) suggests a natural generalization for the
irreducible contact terms, which we have verified at the next order ($k=4$)
\begin{equation}
C\(\phi_{n_1},\ldots,\phi_{n_k}\) =
-{\cal D}^{k-1} (\phi_{n_1}\cdots\phi_{n_k})_-
\label{conttt}
\end{equation}
where
\begin{equation}
{\cal D} f(x) = \left( {f\over W'} \right)'\stop
\label{cald}
\end{equation}
Note that for positive $n_1,\ldots,n_{k-4}$ (\ref{conttt}) is easily
proved by induction. The natural generalization of (\ref{contt}) for $k-1$
derivatives consists of all possible reducible contacts to this order
together with (\ref{conttt}).
The structure of multicontacts that emerges is parity invariant,
although part of the tachyons are primaries and part are descendants.
This is of course expected from the $2d$ string viewpoint, but is
highly non-trivial feature of the topological structure.
The existence of irreducible contact terms of arbitrarily large
numbers of fields (in contrast to the $k>0$ case in which a finite
set yields all contact contributions by iteration as in the first
terms in (\ref{contt})) is the reflection in this language of the
nonlinearity of the recursion relations.

\section{Topological Landau-Ginzburg via periods}

In the preceding sections we have defined our Landau-Ginzburg theory
through (\ref{3}) and (\ref{mpt}). Recently, Eguchi et. al.
\cite{EguchiP} proposed a different definition of the correlators;
they showed that this leads to equivalent results for the
minimal topological models. In this section we will pursue this
approach and show that the defining equations become for $k=-3$
precisely the $W_{1+\infty}$ recursion relations of section two. This
approach does not rely on the existence of a nondegenerate two-point
function between primary fields or on the ring structure they exhibit and
is thus extremely well suited to our problem, where both of these are
absent.

Let us start by reviewing the period integrals approach,
using the $A_{k+1}$ minimal theories as an example. From the known
relation of the Landau-Ginzburg superpotential to the
KdV Lax operator \cite{DVV} Eguchi et. al. extract formulas for one-point
functions of primary fields on the small phase space as periods:
\begin{equation}
\<\<\phi_i(t)\>\> = \frac{(k+2)^{1+\frac{i+1}{k+2}}}{(i+1)(i+k+3)}
\res\( W^{1+\frac{i+1}{k+2}}\)\stop
\label{1pt}
\end{equation}
Integrability by itself is not enough to determine the theory, \ie\
to determine
the superpotential and the correlators. Clearly, the Lax operator
is not unique and therefore the superpotential is not
uniquely defined by specifying the hierarchy.
The additional input is the string equation \cite{Douglas} that implies
\begin{equation}
t_{k-i} = \frac{(k+2)^{\frac{i+1}{k+2}}}{i+1}\res(W^{\frac{i+1}{k+2}})
\comma
\label{fcoor}
\end{equation}
as an equation for the flat coordinates \cite{DVV,EguchiP}.
Equations (\ref{1pt}) and (\ref{fcoor}) determine uniquely the dynamics of the
primary fields.

Using the representation of gravitational descendants by matter
degrees of freedom \cite{Lossev,EguchiD,EguchiP},
one-point functions of descendants on the small phase space
are given as periods:
\begin{equation}
\eqalign{
\<\<\sigma_n(\phi_i)\>\>& = c_{n,i}\,\res(W^{n+1+\frac{i+1}{k+2}})\ ,\cr
c_{n,i} = ((i+1)(i+1+k+2)&\cdots(i+1+(n-1)(k+2))^{-1},\
c_{0,i} = 1\ .\cr
}
\label{d1pt}
\end{equation}

As periods, these satisfy Gauss-Manin equations \cite{EguchiP} which
are equivalent to the recursion relations for the insertion of a
descendant field in a correlator of primaries in topological gravity
\cite{WittTG}. This leads to an alternative formulation of the
Landau-Ginzburg theory. In this approach, the three relations
(\ref{1pt}), (\ref{fcoor}), and (\ref{d1pt}) are taken as the
fundamental defining relations. These lead to a complete and
unambiguous set of correlators, which for all of the minimal models
(and some additional examples) were shown to be identical to those
obtained from the standard Landau-Ginzburg definition of the model.
We note here that correlators with more than one descendant operator
insertion are not directly computable using Landau-Ginzburg techniques
in the minimal models; they are best obtained by using the recursion
relations of \cite{WittTG}.

Two basic ingredients in the traditional approach that are missing in
the topological Landau-Ginzburg description of the $2d$ string are
the ring structure and the metric.
The metric on the space of primary fields
\beq
\eta_{ij} = \<\CT_i\CT_j\CT_1\> \comma
\label{metric}
\eeq
vanishes by momentum conservation, and as a consequence the
ring structure of the primary operators is not computable from the
correlators.

One is led to adopt the definition of topological Landau-Ginzburg
theory through periods for the description of the $2d$ string.
Doing so, one sees that equation (\ref{d1pt}) for $i = -1$ is, up
to a normalization, the recursion relation (\ref{nwi}) for a negative
momentum tachyon $\CT_{-n}$.
Note that while equation (\ref{d1pt}) is only valid
on the small phase space, equation (\ref{nwi}) is valid on a larger
phase space which includes non-negative momentum
tachyons (primary fields) as well as negative momentum tachyons
(descendants).

Equation (\ref{d1pt}) suggests that the negative momentum
tachyons $\CT_{-n}$ are descendants of $\CT_0$ which corresponds to
$\phi_{-1}$.
The latter is not in the BRST cohomology for $k>0$, but exists for $k<0$.
Since the $U(1)$ weights
of all fields are integers, it is not possible to exclude entirely
the possibility that negative momentum tachyons are descendants of some
other primary. However, the proposal that they descend from $\CT_0$ is by
far the most natural.
Note that one can also get the recursion relation for a negative
momentum tachyon $\CT_{1-i}$ from equation (\ref{1pt}) for negative $i$.

Flat coordinates are given by periods with negative powers as in
(\ref{fcoor}). In the $2d$ string:\footnote{For $n=1$ this equation is
indeterminate. In fact the correct resolution is $t_{-1} = \res \log(-W)$.}
\beq
t_{-n} = \frac{1}{1-n} \res(-W)^{1-n}\stop
\label{ncoo}
\eeq

The natural appearance of the $W_{1+\infty}$ recursion relations in this
approach is very encouraging. The difficulty here is that the dynamics
of primary fields in our model is trivial, and the equations (\ref{1pt}),
(\ref{fcoor}), and (\ref{d1pt}) do not by themselves determine the
descendant correlators unambiguously. We remark that it is a property
of the Landau-Ginzburg formulation in general that descendant dynamics
are not as readily expressed as those of the primaries.
However, if we add the additional constraint
\begin{equation}
W = -\frac{1}{x} \(1 + \sum_{n=0}^\infty t_n x^n + {\cal O}(x^{-1})\)
\label{constraint}
\end{equation}
which is manifestly satisfied by (\ref{itW}) then the solution becomes
unique. This enforces the absence of contact terms between primaries
as noted following (\ref{WPS}).

The self-consistency of the periods formulation and the fact that it
fully describes
the tachyon dynamics strongly indicates that our Landau-Ginzburg formulation of
 the
$2d$ string is indeed correct. Certainly, more work is needed in order to
uniquely describe the descendants dynamics in topological
Landau-Ginzburg theory in general, and in the $2d$ string in particular.

\section{ Integrable structure of 2d string theory}

The integrable structure of $c\leq 1$ conformal matter coupled
to gravity, or alternatively minimal topological matter coupled to topological
gravity, reveals itself via the identification of the generating functional
of the theory as a tau function of the corresponding integrable
hierarchy.
In the $c=1$ case the integrable hierarchy is the Toda Lattice
hierarchy \cite{ronen} which includes  KP,
generalized KdV and various other hierarchies.

In Landau-Ginzburg theories the integrable structure appears in its Lax
formulation.
In the $A_{k+1}$ models the Landau-Ginzburg superpotential is identified with
 the
Lax operator of the corresponding $k^{th}$ generalized KdV hierarchy
\cite{DVV}.
As we will see the above identification is modified in the
$c=1$ case, and the superpotential is related to the Baker-Akhiezer
wave function.

\subsection{Toda lattice hierarchy}
In this section we introduce the Toda lattice hierarchy, which will
be used in the next section in order to establish the connection between
the topological Landau-Ginzburg formulation and the integrable
structure of the $2d$ string.

The Toda Lattice hierarchy \cite{Ueno}
is a generalization of the KP hierarchy and
includes two sets of times $t=(t_1,t_2,\ldots), \bar{t}=(\bar{t}_1,
\bar{t}_2,\ldots)$ and an additional parameter $s$ which we will identify
respectively with the couplings to positive and negative tachyons and
the cosmological constant $1+t_0$. One introduces two Lax operators
\begin{equation}
\eqalign{
L&= \sum_{j=-1}^{\infty}a_j(s,t,\bar{t},\hbar)D^{-j}\cr
M&= \sum_{j=-1}^{\infty}b_j(s,t,\bar{t},\hbar)D^j  ,\cr
}
\label{Laxop}
\end{equation}
where $D=e^{\hbar\partial_s}$ is a shift operator.
The limit $\hbar\rightarrow 0$ yields
the dispersionless limit of the hierarchy which in the two dimensional
string language corresponds to the restriction to genus zero.
Note that unlike the KP hierarchy $D$ is not a pseudo differential operator.
In fact the naive attempt to represent the
Lax formulation of the hierarchy using pseudo differential operators
by replacing $D\rightarrow \pa_x$ leads
to inconsistencies.

The Lax type equations are
\begin{equation}
\eqalign{
\hbar\partial_n L &= \[L_+^n, L\]      \cr
\hbar\partial_n M &= \[L_+^n, M\]      \cr
\hbar\bar{\partial}_n L &= \[M_-^n, L\] \cr
\hbar\bar{\partial}_n M &= \[M_-^n, M\]  \ ,\cr
}
\label{Laxeq}
\end{equation}
where $\partial_n = \partial_{t_n}, \bar{\partial}_n = \partial_{\bar{t}_n}$.

A connection between the Lax operators and the tau function is
established as follows. We introduce two diagonalizing operators
\begin{equation}
\eqalign{
\Omega &= 1 + \sum_{j=1}^{\infty} \Omega_j(s,t,\bar{t},\hbar)D^{-j}\cr
\Lambda &= \sum_{j=0}^{\infty} \Lambda_j(s,t,\bar{t},\hbar) D^j\ ,\cr
}
\label{diagop}
\end{equation}
and their inverse operators
\begin{equation}
\eqalign{
\Omega^{-1} &= 1 + \sum_{j=1}^{\infty} D^{-j}\Omega_j^{\ast},
(s+\hbar,t,\bar{t},\hbar)\cr
\Lambda^{-1} &= \sum_{j=0}^{\infty}
D^j\Lambda_j^{\ast}(s+\hbar,t,\bar{t},\hbar)\ ,\cr
}
\label{invop}
\end{equation}
such that
\begin{equation}
\eqalign{
L &= \Omega D \Omega^{-1},       \cr
M &= \Lambda D^{-1} \Lambda^{-1} \ .\cr
}
\label{Laxdiag}
\end{equation}
The Lax equations can be converted to the Sato equations
\begin{equation}
\eqalign{
\hbar\partial_n \Omega &= -L_-^n \Omega \cr
\hbar\partial_n \Lambda &= L_+^n\Lambda \cr
\hbar\bar{\partial}_n \Omega &= M_-^n \Omega \cr
\hbar\bar{\partial}_n \Lambda &=-M_+^n\Lambda \ .\cr
}
\label{sato}
\end{equation}

The operators $\Omega, \Lambda$ and their inverse are related to the
tau function via
\begin{equation}
\eqalign{
\sum_{j=0}^{\infty} \Omega_j(s,t,\bar{t},\hbar) k^{-j} &=
{\tau(s,t-\hbar\varepsilon(k^{-1}),\bar{t},\hbar)\over
\tau(s,t,\bar{t},\hbar)}\cr
\sum_{j=0}^{\infty} \Lambda_j(s,t,\bar{t},\hbar) k^{j}
&={\tau(s+\hbar,t,\bar{t}-\hbar\varepsilon(k),\hbar)\over
\tau(s,t,\bar{t},\hbar)}\cr
\sum_{j=0}^{\infty} \Omega_j^{\ast}(s,t,\bar{t},\hbar) k^{-j} &=
{\tau\(s,t+\hbar\varepsilon\(k^{-1}\),\bar{t},\hbar\)\over
\tau(s,t,\bar{t},\hbar)}\cr
\sum_{j=0}^{\infty} \Lambda_j^{\ast}(s,t,\bar{t},\hbar) k^{j}
&={\tau(s-\hbar,t,\bar{t}+\hbar\varepsilon(k),\hbar)\over
\tau(s,t,\bar{t},\hbar)} \ ,\cr
}
\label{tau}
\end{equation}
where $\varepsilon(k) = \(k, k^2/2,k^3/3,\dots\)$.
Equations (\ref{Laxdiag}), (\ref{tau}) relate the Lax operators and the
$\tau$ function.

One has \cite{ronen}
\begin{equation}
\log \tau (\hbar,t,\bar{t},s) = \hbar^{-2} F(t,\bar{t},s)
\comma
\label{logtau}
\end{equation}
with $F$ the generating functional of the two dimensional
string, and the genus expansion corresponding to an expansion in $\hbar^2$.
Evidently, this is a specific Toda lattice $\tau$ function, corresponding to
picking
a point in the Sato grassmanian. This point is determined by the $2d$ string
$S$-matrix calculation of \cite{exacts}.

The relations (\ref{tau}) can be written more compactly as
\begin{equation}
\eqalign{
\Omega_j(s,t,\bar{t},\hbar) &= \frac{1}{\tau(s,t,\bar{t},\hbar)}
P_j\(-{\cal D}\) \tau(s,t,\bar{t},\hbar)\cr
\Lambda_j(s,t,\bar{t},\hbar) &= \frac{1}{\tau\(s,t,\bar{t},\hbar\)}
P_j\(-\bar{\cal D}\) \tau(s+\hbar,t,\bar{t},\hbar) \ ,\cr
}
\label{Schur}
\end{equation}
where $P_j$ is the $j$th Schur polynomial defined by
\beq
\exp\[\sum_{n=1}^{\infty}x_n\lambda^n\] = \sum_{n=0}^{\infty}P_n(x)\lambda^n
\comma
\label{Schu}
\eeq
and ${\cal D} = \(\pa_1,\pa_2/2,\pa_3/3,\ldots\)$.

Finally, we introduce the Baker-Akhiezer functions
\begin{equation}
\eqalign{
\Psi &= \(\sum_{j=0}^{\infty} \Omega_j\(s,t,\bar{t},\hbar\) k^{-j}\)
\exp {1\over\hbar} \[\sum_{n=1}^{\infty}t_nk^n + s \log k\]\cr
\Phi &=\(\sum_{j=0}^{\infty} \Lambda_j\(s,t,\bar{t},\hbar\) k^j\)
\exp {1\over\hbar} \[\sum_{n=1}^{\infty}\bar{t}_nk^{-n} + s \log k\]\ .\cr
}
\label{BA}
\end{equation}
$\Psi$, $\Phi$ are eigenfunctions of the Lax operators $L$ and $M$
respectively.

\subsection{Relating the integrable and topological Landau-Ginzburg
structures}
In this section we unravel the connection between the topological
Landau-Ginzburg
and the integrable structures of the $2d$ string.
Since the Landau-Ginzburg description is restricted to genus zero, we take
the dispersionless limit of the Toda lattice hierarchy (following
\cite{TakTak}) \ie\ $\hbar\rightarrow 0$.
This is done by replacing $D\rightarrow p$,
$[\cdot,\cdot]\rightarrow\{\cdot,\cdot\}$ where the Poisson
brackets are defined on the phase space $(p,s)$ as
\begin{equation}
\{f(p,s),g(p,s)\} = p\partial_p f \partial_s g - \partial_s f p\partial_p g
\stop
\label{pb}
\end{equation}

The Baker-Akhiezer functions, which satisfy time-dependent Schr\"odinger-like
equations take the WKB form
\begin{equation}
\eqalign{
\Psi &= \exp\[{1\over\hbar} S(t,\bar{t},s,k) + O\(\hbar^0\)\]\cr
\Phi &= \exp\[{1\over\hbar} \bar{S}(t,\bar{t},s,k) + O\(\hbar^0\)\]\ \cr
}
\label{WKB}
\end{equation}

Plugging (\ref{tau}) in (\ref{BA}) and comparing to (\ref{WKB}) using
(\ref{logtau})
we see that the superpotential $W$ in equation (\ref{SW}) and its parity dual
in equation (\ref{barW}) $\bar{W}$ are
identified as
\begin{equation}
W(k) = -\partial_k S(k),\qquad \bar{W}(k) = -\partial_k \bar{S}(k) \comma
\label{WBA}
\end{equation}
and $s = 1 +t_0$.
This is the first main result of this section.
It establishes the connection between the Landau-Ginzburg formalism
and the
Toda hierarchy, and parallels the identification of the Landau-Ginzburg
superpotential
$W$ in the $A_{k+1}$ model for $k>0$ with the Lax operator of the
$k^{th}$ KdV hierarchy \cite{DVV}.

This provides a natural guess for the generalization of
the superpotential to higher genus, that is
\begin{equation}
W = - \hbar \partial_k \log \Psi \stop
\label{genW}
\end{equation}

Let us now relax the restriction to genus zero.
Using (\ref{Schur}) and the Sato equations (\ref{sato}) we get
\begin{equation}
\eqalign{
\hbar^2\partial_1\partial_n\log\tau &= \res\(L^n\)\cr
\hbar^2\partial_1\bar{\partial}_n\log\tau &= -\res\(M^n\) \ .\cr
}
\label{resL}
\end{equation}
Upon identification of the generating functional of the $c=1$ theory
with the tau function of the Toda Lattice hierarchy we get
\begin{equation}
\eqalign{
\res\(L^n\) &= \<\<\CT_1\CT_n\>\>,       \cr
\res\(M^n\) &= - \<\<\CT_1\CT_{-n}\>\>\ .\cr
}
\label{resLn}
\end{equation}
This is the second main result of this section.
It implies that the tachyon with momentum one $\CT_1$ is
indeed the puncture operator in the topological description of the $2d$
string, thus confirming our identification.
Analogous relations for the case of minimal conformal theories coupled
to gravity are derived by integrating the Gelfand-Dikii equations using
the string equation \cite{Wittalgeom}.

The parity transformed relations are
\begin{equation}
\eqalign{
L^n_1 &= -\<\<\CT_{-1}\CT_n\>\>\cr
M^n_1 &= \<\<\CT_{-1}\CT_{-n}\>\>\ ,\cr
}
\label{Lnone}
\end{equation}
where the subscript $1$ denotes the coefficient of $D$.
In fact, parity invariance of a theory with
the Toda lattice as its underlying integrable structure
implies that $L$ determines $M$ and vice versa.

Similar relations hold also for the insertion of the
cosmological operator $\CT_0$, we have
\begin{equation}
\eqalign{
L^n_0&=\<\<T_0T_n\>\> \cr
M^n_0&=-\<\<T_0T_-n\>\>\ ,\cr
}
\label{Lnzero}
\end{equation}
where the subscript $0$ denotes the coefficient of the identity.
Note that in the minimal models, the $0$ coefficient of the Lax operator
 vanishes,
which is consistent with the fact that the analogue of this operator is BRST
 exact.

\section{Conclusions}

We have, in this paper, constructed a topological Landau-Ginzburg theory
which computes the tachyon scattering matrix in $2d$ string theory
at the self-dual radius. The most glaring weakness of the approach is
the lack of an understanding of the worldsheet topological field theory
we are describing. The nonpolynomial nature of the superpotential poses
a serious obstacle to such understanding. However, even in its present
state, the Landau-Ginzburg formulation is a useful representation of
the structure of the contact term algebra in the theory. We have uncovered
an extremely rich structure reflecting the intricate momentum
dependence  of the
scattering amplitudes. Surprisingly, we were able to write the
Landau-Ginzburg superpotential even away from the small phase space,
exhibiting explicitly the dependence on the descendant ``times''.
This gives a formulation of the theory
that is manifestly background independent (of course at present this
refers exclusively to tachyon backgrounds;). This can hopefully be
used, for example, to study the phase structure of the theory along
the lines of \cite{SG}.

In recent work
\cite{gkm} a relation between the Landau-Ginzburg superpotential and
the potential of the ``generalized Kontsevich model'' was discovered.
We note that restricting attention to the small phase space our results
satisfy this relation to the potential of the Kontsevich-like
representation for $2d$ string amplitudes found in \cite{ronen}. An
extension of this relation beyond the small phase space is not known but
since we have found the superpotential it is natural to conjecture that
the same relation will hold throughout the parameter space.

We were led by this formulation to an interpretation of part of the
spectrum of states: positive momentum tachyons are the primary fields,
negative momentum tachyons are descendants, probably of the cosmological
constant operator $\CT_0$.
$\CT_1$ is the puncture operator and its recursion relation (\ref{T1WI})
is the
puncture equation. This interpretation is supported by the topological
formulation of the model following \cite{WittenNM}. Inserting $\CT_k$ into
a correlator modifies the definition of the bundle $\cal V$ by tensoring
with ${\cal O}(z)^{1-k}$. For $k=1$ this marks a point, as one expects from
a puncture.
Note that in contrast to the minimal models, in which the cosmological
constant couples to the dilaton operator, in the $2d$ string it couples
to a primary $\CT_0$. The peculiar property of $\CT_0$ is that it has
the same $U(1)$ weight as the superpotential itself; the minimal models
do not contain such a primary operator.

The natural question that arises is what observables of the $2d$
string correspond to the gravitational descendants of the other primary
fields; the obvious answer is the discrete states. Our description of the
tachyon spectrum generalizes as
\beq
\CY^+_{j,m} = \sigma_k(\CT_n) \ ,
\label{discid}
\eeq
with $k = j-m$ and $n = j+m$.
Figure one shows the spectrum of $2d$ string theory and its proposed
structure as a topological gravity theory.

\begin{figure}[h]
\setlength{\unitlength}{0.0125in}%
\begin{picture}(450,370)(100,400)
\put(331,440){\makebox(0,0)[lb]{\raisebox{0pt}[0pt][0pt]{${\cal T}_0$}}}
\put(309,550){\makebox(0,0)[lb]{\raisebox{0pt}[0pt][0pt]{${\cal Y}^+_{1,0}$}}}
\put(252,601){\makebox(0,0)[lb]{\raisebox{0pt}[0pt][0pt]
{${\cal Y}^+_{3/2,-1/2}$}}}
\thicklines
\put(340,460){\vector(-1, 1){ 60}}
\put(280,520){\vector(-1, 1){ 60}}
\put(220,580){\vector(-1, 1){ 60}}
\put(160,640){\vector(-1, 1){ 60}}
\put(340,460){\line( 1, 1){240}}
\put(400,520){\vector(-1, 1){ 60}}
\put(340,580){\vector(-1, 1){ 60}}
\put(280,640){\vector(-1, 1){ 60}}
\put(460,580){\vector(-1, 1){ 60}}
\put(400,640){\vector(-1, 1){ 60}}
\put(520,640){\vector(-1, 1){ 60}}
\put(267,491){\makebox(0,0)[lb]{\raisebox{0pt}[0pt][0pt]{${\cal T}_{-1}$}}}
\put(209,548){\makebox(0,0)[lb]{\raisebox{0pt}[0pt][0pt]{${\cal T}_{-2}$}}}
\put(152,606){\makebox(0,0)[lb]{\raisebox{0pt}[0pt][0pt]{${\cal T}_{-3}$}}}
\put(87,672){\makebox(0,0)[lb]{\raisebox{0pt}[0pt][0pt]{${\cal T}_{-4}$}}}
\put(398,497){\makebox(0,0)[lb]{\raisebox{0pt}[0pt][0pt]{${\cal T}_1$}}}
\put(455,553){\makebox(0,0)[lb]{\raisebox{0pt}[0pt][0pt]{${\cal T}_2$}}}
\put(515,613){\makebox(0,0)[lb]{\raisebox{0pt}[0pt][0pt]{${\cal T}_3$}}}
\put(576,675){\makebox(0,0)[lb]{\raisebox{0pt}[0pt][0pt]{${\cal T}_4$}}}
\put(198,671){\makebox(0,0)[lb]{\raisebox{0pt}[0pt][0pt]{${\cal Y}^+_{2,-1}$}}}
\put(376,609){\makebox(0,0)[lb]{\raisebox{0pt}[0pt][0pt]
{${\cal Y}^+_{3/2,1/2}$}}}
\put(313,672){\makebox(0,0)[lb]{\raisebox{0pt}[0pt][0pt]{${\cal Y}^+_{2,0}$}}}
\put(433,673){\makebox(0,0)[lb]{\raisebox{0pt}[0pt][0pt]{${\cal Y}^+_{2,1}$}}}
\end{picture}

Figure 1 -- {\it The spectrum of $2d$ string theory as a theory of topological
gravity. The horizontal axis is the momentum, the vertical axis the
energy. Arrows connect a state to its descendant.}
\vskip .2in
\end{figure}
In the spirit of the period integrals (\ref{d1pt}),
one may conjecture their one point function
on the space of tachyon backgrounds:
\beq
\<\<\CY_{j,m}^+\>\>= c_{j,m}\res\(-W\)^{1-2m}
\qquad -j \leq m < 0 \ ,
\label{ds1pt}
\eeq
for some constants $c_{j,m}$. One point functions of $\CY^+_{j,m}$
for positive $m$ are the parity transformed versions of (\ref{ds1pt}).
It would be extremely interesting to verify this conjecture and
compute the correlators of these states. Despite much effort,
there is as yet no reliable calculation of their dynamics for
$\mu\neq 0$.
Note that we do not expect that special states and tachyons  will appear
on an equal footing in the full
description of the theory. This is similar to the asymmetric appearance of
primaries and descendants in minimal models. However, in view of the
symmetry structure of the $2d$ string, we expect that $\CY_{j,m}^+$ and
$\CY_{j,-m}^+$ will appear symmetrically much as the negative and positive
tachyons do.

The topological model we have constructed differs from
the traditional topological theories. First, it has an infinite number
of gravitational primaries. Second, the metric on the space of primaries
vanishes. Thus a naive extrapolation of Witten's recursion relations is
not applicable. An understanding of the recursion relations from the
point of view of two-dimensional topological gravity could enrich our
understanding of this theory. From the topological Landau-Ginzburg
point of view it
is different in that the standard algebraic manipulations fail to
generate the ring structure on the space of primaries.
It turned out, however, that the definition of topological Landau-Ginzburg
through period integrals is very natural in this case, which indicates
that there is a class of topological field theories for which the
traditional Landau-Ginzburg theory generalizes.

An extremely interesting open problem seems to be the generalization of the
topological description of the $2d$ string to higher
genus. Two approaches to accomplish this goal seem to appear from our
work. The first is to try to generalize the period integral definition
to higher genus using the $W_{1+\infty}$ identities. As we saw,
the genus zero restriction of these emerged very naturally in our work.
One may hope that a similarly natural role for the general case could
be found.

The second approach to the problem is through the integrable structure
of the theory. As we saw there is a natural candidate for the superpotential
at higher genus (\ref{genW}). Moreover, the KdV integrable structure
of the minimal models turned out to be a very powerful computing tool.
Of particular importance is the string equation. The string equation is
not known in the $c=1$ case, and it would be highly desirable to find it.
The Toda lattice hierarchy provides a natural guess: Consider the operator
\beq
{\cal
 M}=-\sum_{n=1}^{\infty}t_nL^{n-1}-sL^{-1}-
\sum_{n=1}^{\infty}nv_n(t,\bar{t},s,\hbar)L^{-n-1}\comma
\label{MS}
\eeq
where $L$ is the Toda lattice Lax operator.
In the dispersionless limit ${\cal M}$ acting on the Baker-Akhiezer
wave function
coincides with the Landau-Ginzburg superpotential, i.e. $W = {\cal
 M}(L\rightarrow x)$,
with
$v_n \equiv  \hbar^2 \pa_n \log \tau$, where $\tau$ is the Toda lattice tau
 function.
Upon identification
of $\log \tau$ with the $2d$ string generating functional, $v_n$ is the one
 point function
$\<\<T_n\>\>$. Its expansion in powers of $\hbar^2$ corresponds to genus
 expansion.
One has the equation \cite{TakTak}
\beq
\[{\cal M},L\] = 1 \stop
\label{stringeq}
\eeq
This seems to be the analog of the string equation \cite{Douglas}
\beq
\[P,Q\] = 1 \comma
\label{stringreq}
\eeq
in the minimal models.

Note that in contrast to the KdV hierarchy
which is naturally described using pseudo-differential operators,
the Toda lattice hierarchy is described by shift operators. This leads to
some extra complications, but
the structure seems to parallel that of the KdV case, see for
instance equation (\ref{resLn}).

\noindent{\it Note Added}

A rather similar proposal for a topological Landau-Ginzburg
formulation of $2d$ string theory has recently been made in
\cite{Ghoshal}. The notable difference between the two papers
is that we obtain a Landau-Ginzburg formulation valid beyond
the small phase space.

After submitting this work we received two preprints in which the
integrable structure of section five was further pursued
\cite{Takasaki,Eguchii}. This work leads to a proof of
the conjecture formulated in section two as follows.
Recall that the system of equations (\ref{3}),(\ref{mpt})
and (\ref{itW}) recursively determines the correlators and hence
the superpotential. To prove that the Landau-Ginzburg formulation indeed
reproduces the correlators of $2d$ string theory it suffices to show
that the latter satisfy (\ref{3}) considered as a recursion relation
with $W,\phi$ on the right-hand side given by (\ref{SW}),(\ref{Phi})
in terms of $2d$ string theory correlators.

In \cite{Takasaki,Eguchii} a particular solution to the Toda lattice hierarchy
was constructed by imposing a string equation-like condition on the Lax
operators. The solution to this was shown to be unique and to satisfy
(\ref{pwi}),(\ref{nwi}), hence the correlators it computes are those of
the $2d$ string. Defining $W,\bar W$ in terms of these the string equation
reads\footnote{We use $\bar x$ for the argument of $\bar W$ here, to retain
clarity when both functions appear together.}
\begin{equation}
\eqalign{
x &= -\bar{W}(-W(x))\cr
\bar{x} &= -W(-\bar{W}(\bar{x}))\ .\cr}
\label{String}
\end{equation}
In terms of $W$ eqn. (17) of \cite{Takasaki} is
\beq
{1\over m} x^m - \sum_{n>0} \<\< T_m T_n\>\> x^{-n} =
\<\< T_m T_0\>\> + \sum_{n>0}\<\< T_m T_{-n}\>\> (-W)^{-n}\ .
\label{Tident}
\eeq
Taking the $x$ derivative and comparing (\ref{Phi}) we have
\beq
\phi_m = \sum_{n>0} n \<\< T_m T_{-n}\>\> (-W)^{-n-1} W'
\label{Tphi}
\eeq
where for simplicity we assume $m>0$.
Taking the $t_k$ derivative of (\ref{Tident}) and using (\ref{Tphi})
we find
\beq
\sum_{n\geq 0} \<\< T_k T_m T_n\>\> x^{-n} =
{\phi_k \phi_n\over W'}-
\sum_{n>0} \<\< T_k T_m T_{-n}\>\> (-W)^{-n} \ .
\label{almost}
\eeq
We now multiply (\ref{almost}) by $\phi_{-l}$ and evaluate the residue.
The left-hand side vanishes. On the right-hand side we can use (\ref{String})
to evaluate\footnote{This manipulation also verifies the conjecture
(\ref{ncoo}).}
\beq
{\rm res}_x \left( \phi_{-l}(x) (-W(x))^{-n} \right) =
{1\over 1-n} \partial_{-l}{\rm res}_{\bar x}
\left( \bar{x}^{1-n} \bar{W}'\right)
 = \delta_{n,l}\ ,
\label{trick}
\eeq
leading precisely to (\ref{3}).

\appendix
\section{Verifying the recursion relations}
In this appendix we continue the program started in section two of
computing correlators order by order in the deviation from the small
phase space and verifying that the recursion relation (\ref{nwi})
holds. As in that section, all final results understood to be
evaluated on the small phase space.

The next prediction one can test following from (\ref{nwi}) is
obtained from (\ref{twomin}) by taking the derivative
\begin{equation}
\<\<\CT_{-n_1}\CT_{-n_2}\CT_{-n_3}\>\> =
\res \left\{ \phi_{-n_2}\phi_{-n_3} (-W)^{n_1-1} -
\partial_{-n_3} \phi_{-n_2} \[-\frac{1}{n_1}(-W)^{n_1}\]\right\}\ .
\label{threemin}
\end{equation}
One can further use (\ref{Phi}) to find
\beq
\partial_{-n_2}\phi_{-n_3} = \sum_{m>0} mx^{-m-1}
\res \left( {x^{m-1}\phi_{-n_2}\phi_{-n_3}\over W'}\right)
= -\left( {\phi_{-n2}\phi_{-n3}\over W'} \right)'_- \ ,
\label{mimi}
\eeq
verifying (\ref{cont}). Plugging this into (\ref{threemin}) and using
(\ref{useful}) we obtain
\begin{equation}
\eqalign{
\<\<\CT_{-n_1}\CT_{-n_2}\CT_{-n_3}\>\> &=
\res \left\{ {\phi_{-n_2}\phi_{-n_3}\over W'} \(
 (-W)^{n_1-1}W' +\frac{1}{n_1}\[(-W)^{n_1}\]'_+
\)\right\}\cr
&= \res \( { \phi_{-n_1}\phi_{-n_2}\phi_{-n_3}\over W'}\)\ ,\cr
}
\label{threemini}
\end{equation}
verifying agreement to second order.

We wish to extend this argument to third order. This is not merely an
attempt to push the calculation, but is actually important in two
ways. First, this is the order at which computing (\ref{3}) will
require information not available from study of the small phase space
only. In process of computing to this order, we will also find the
first of the multipoint contact terms referred to in the text.
Taking the derivative of (\ref{threemin}) we find
\begin{equation}
\eqalign{
\<\<\CT_{-n_1}\CT_{-n_2}\CT_{-n_3}\CT_{-n_4}\>\>  =
\res &\Biggl\{ \phi_{-n_2}\phi_{-n_3} \partial_{-n4} (-W)^{n_1-1} +
\partial_{-n_4}(\phi_{-n_2}\phi_{-n_3}) (-W)^{n_1-1} \cr
& + \phi_{-n_4} \partial_{-n_3} \phi_{-n_2} (-W)^{n_1-1} -
 \partial_{-n_4}\partial_{-n_3}\phi_{-n_2} \[-\frac{1}{n_1}(-W)^{n_1}\]
\Biggr\} \ .\cr
}
\label{fourmin}
\end{equation}
To proceed we get from (\ref{Phi})
\begin{equation}
\partial_{-n_4}\partial_{-n_3}\phi_{-n_2} =
-\left( \sum_{m>0} x^{-m} \<\<\CT_m\CT_{-n_2}\CT_{-n_3}\CT_{-n_4}\>\>
\right)'_-
\label{twoders}
\end{equation}
where we note that now the correlators in the sum fall under those for
which we have proved agreement above. We thus evaluate
\begin{equation}
\eqalign{
&\<\<\CT_m\CT_{-n_2}\CT_{-n_3}\CT_{-n_4}\>\>
 =\partial_{-n_4}\res \({\phi_m\phi_{-n_2}\phi_{-n_3}\over W'}\)\cr
& =\res\left\{ \partial_{-n_4}(\phi_{-n_2}\phi_{-n_3}){x^{m-1}\over W'}
- {\phi_{-n_2}\phi_{-n_3}\over W'}
\[ \( {x^{m-1}\phi_{-n_4}\over W'}\)'_--{x^{m-1}\over W'}\phi_{-n_4}'
\]\right\}\cr
&=\res\left\{ {x^{m-1}\over W'} \[
\phi_{-n_2}\partial_{-n_4}\phi_{-n_3} +
\phi_{-n_3}\partial_{-n_4}\phi_{-n_2} +
\phi_{-n_4}\partial_{-n_3}\phi_{-n_2} +
\left({\phi_{-n_2}\phi_{-n_3}\phi_{-n_4}\over W'}\right)'
\] \right\} \cr
}
\label{compute}
\end{equation}
where we hope the reader can complete the omitted steps. Inserting in
(\ref{twoders}) now yields the result quoted in the text as
(\ref{contt})
for the multipoint contact terms.

Finally, one then inserts this, in turn, in (\ref{fourmin}) and finds
\begin{equation}
\eqalign{
&\<\<\CT_{-n_1}\CT_{-n_2}\CT_{-n_3}\CT_{-n_4}\>\> =
\res\left\{
\partial_{-n_4}(\phi_{-n_2}\phi_{-n_3}){\phi_{-n_1}\over W'}  +
\partial{-n_3}\phi_{-n_2} {\phi_{-n_1}\phi_{-n_4}\over W'} + \right.\cr
&\qquad \left.\phi_{-n_2}\phi_{-n_3}\partial_{-n_4}(-W)^{n_1-1} -
{1\over W'} {\phi_{-n_2}\phi_{-n_3}\phi_{-n_4}\over W'}
\[-\frac{1}{n_1} (-W)^{n_1}\]_+ \right\}\cr
&= \res\left\{
\partial_{-n_4}(\phi_{-n_2}\phi_{-n_3}){\phi_{-n_1}\over W'} +
 {\phi_{-n_2}\phi_{-n_3}\over W'}\partial_{-n_4}\phi_{-n_1} +
{\phi_{-n_1}\phi_{-n_2}\phi_{-n_3}\over (W')^2} \phi_{-n_4}'
\right\}\cr
}
\label{compare}
\end{equation}
in agreement with the prediction of (\ref{mpt}) verifying the agreement
to third order.

\vskip 1 cm

{\large \bf \noindent Acknowledgments}
It is a pleasure to thank
R. Dijkgraaf, T.~Eguchi, K. Intriligator, D.~Kutasov, A. Lossev,
A.~Schwimmer, C.~Vafa, and especially G.~Moore for helpful
discussions, and G. Moore and S. Shatashvili for comments on the manuscript.
R.P. thanks the department of physics at the Weizmann Institute, where
much of this work was completed, for their hospitality.

\end{document}